\documentclass
[twocolumn,aps,prc,amsmath,showpacs,amssymb,floatfix]
{revtex4}
\usepackage{amssymb}

\usepackage{CJK}                     
\usepackage[dvips]{graphicx}
\usepackage{bm}                      
\usepackage{dcolumn}                 
\usepackage{array}
\usepackage{graphicx}
\usepackage{dcolumn}
\usepackage{bm}
\usepackage{ulem} 
\usepackage[usenames]{color}
\usepackage{epstopdf}
\usepackage{epsfig}
\usepackage{float}
\usepackage{subfigure}
\usepackage{multirow}
\usepackage{slashed}
\usepackage{diagbox}
\usepackage{cancel}
\usepackage{graphics}
\usepackage{longtable,pdflscape}

\newcommand{\nc}{\newcommand}       
\nc{\vc}[1] {\mbox{\boldmath $#1$}} 
\nc{\del}       {\partial}              
\nc{\bra}       {\langle}               
\nc{\ket}       {\rangle}               
\nc{\bras}[1]   {\langle #1|}           
\nc{\kets}[1]   {|#1\rangle}            
\nc{\mapleft}[1]{           
 \smash{\mathop{\,          %
  \hbox to 1.5cm{\rightarrowfill}\, }\limits_{#1}}}
\nc{\beq}     {\begin{eqnarray}}
\nc{\eeq}    {\end{eqnarray}}
\nc{\nn}      {\\\nonumber} \nc{\vs}      {\vspace{-0.275cm}}
\nc{\fra}    {\frac{1}{2}}
\nc{\mb}        {\mathbf}
\usepackage{color}

\begin{document}

\title{Quark mean-field model for nuclear matter with or without bag}
\author{Zhen-Yu Zhu$^{1}$\footnote{zhenyu19921204@live.com}, Ang Li$^{1}$\footnote{liang@xmu.edu.cn}, Jin-Niu Hu$^{2}$\footnote{hujinniu@nankai.edu.cn}, Hong Shen$^{2}$\footnote{songtc@nankai.edu.cn}}
\affiliation{
$^1$Department of Astronomy, Xiamen University, Xiamen 361005, China\\
$^2$School of Physics, Nankai University, Tianjin 300071, China\\
}
\date{\today}

\begin{abstract}
We propose the new quark mean-field bag (QMFB) model by incorporating the bag confinement mechanism in the original quark mean-field model. Nuclear matter and neutron star properties are studied with the QMFB model. For the study of the bag effect, we newly fit 12 parameter sets by reproducing the empirical saturation properties of nuclear matter. Quark confinement is found to be mainly demonstrated by the bag after it is included in the model, instead of the confining potential. For nuclear matter, the bag decreases the binding energy and increases the symmetry energy. For neutron star, the bag affects significantly the radius $R$ of a $1.4M_\odot$ star, with the maximum mass only slightly modified. The bag also has a large suppression effect on the well-accepted $R$ vs $L$ dependence, with $L$ the symmetry energy slope at the saturation density.

\end{abstract}

\pacs{21.60.Jz, 21.65.Cd, 21.65.Mn, 26.60.-c, 97.60.Jd}

\maketitle

\section{Introduction}

The properties of nucleon in nuclear medium may be different with those in free space through the implication of the EMC effect, although other interpretations such as the effects the pion enhancement are not excluded~\cite{emc1,emc2}. Recently, the effective influence on the nuclear matter and neutron star properties from the nucleon radius in free space has been discussed~\cite{pro1,pro2,rN}. Therefore it is important to study from the quark level, by which the properties of nucleons in nuclear medium can be obtained self-consistently.

In 1988, Guichon proposed the quark meson coupling (QMC) model to study nuclear matter~\cite{qmcgui}. This model mimics the relativistic mean-field theory, which describes the nuclear interaction as the exchanges of scalar and vector mesons. But in QMC, the meson fields couple with quarks instead of nucleons, where the properties of nucleon vary from free space in nuclear medium. The nucleon internal structure is described by the MIT bag model, where three quarks have a current quark mass $\sim 5$ MeV and are confined by a spherical bag. This model has been extended to study finite nuclei, hypernuclei, and neutron star in the following years~\cite{qmc1,qmc2,qmc3,qmc4}.

On the other hand, a new interesting model was built by Toki and his collaborators~\cite{tokiqmf,shenqmf}, where the quarks are confined not by the bag but by a central harmonic oscillator potential. The quarks in nucleon are described as constituent quarks, which acquire masses from the spontaneous chiral symmetry breaking. The model is referred to as quark mean-field (QMF) model and has also been extended and applied to several observables of nuclear many-body systems~\cite{rN,wangp1,wangp2,qmf1,qmf2}.

The confinement mechanism of a harmonic oscillator potential was also proposed in the QMC model~\cite{poten1,poten2,poten3}, with the quark mass in these works ranging from $40$ MeV to $5000$ MeV. Furthermore, recent works from Barik and his collaborators developed the modified QMC model, where the center-of-mass ($c.m.$) corrections, the pionic and gluonic corrections were taken into account and three parameter sets with quark mass of $40$, $50$, $300$ MeV were fitted~\cite{panda1,panda2,panda3}. Meanwhile the same model was studied with constituent quark masses of $250$, $300$, and $350$ MeV in the name of QMF model~\cite{humq1,humq2,humq3}. Therefore, the criterion of QMF and QMC model becomes obscured.

In the present work, a new model referred to as the quark mean-field bag (QMFB) model is proposed, by taking both bag and potential into account. It can be regarded as a hybrid framework of QMF which confines constituent quarks with a potential and QMC which confines current quarks with a bag. The quarks in QMFB are considered as constituent quarks and possess a mass close to $300$ MeV. The central potential can be understood as a mean-field treatment of the interaction between quarks, accounting for the interactions and spin correlations between quarks. The properties of nuclear matter and neutron star will be studied in the present QMFB model and the bag effects are explored in details by comparing the results with those of the original QMF model.

The paper is organized as follows. We provide the details of the QMF formula with or without bag in Sec.~II. The properties of nuclear matter and the parameters of meson coupling are given in Sec.~III, before the investigation of the neutron star matter in Sec.~IV. Finally in Sec.~V is the summary of this work.

\section{quark mean-field model}

In QMF, we first construct a nucleon system with three confined quarks before studying the many-body problems. Since the discussion will be done for the comparisons of QMF model with and without bag and the impact of the confining bag, the details of both models are given for completeness.

\subsection{QMF without bag}

The QMF model without bag originates from the work of Refs.~\cite{panda1,panda2,panda3}, then is further used to study finite nuclei and hypernuclei~\cite{humq1,humq2,humq3}. The confinement of this model is achieved by a  central potential with a harmonic oscillator form, and the quarks in nucleons are thought as the constitute ones, whose masses are about $300$ MeV.

We start with the Dirac equation of the confined quarks:
\begin{eqnarray}
\bigg[i\gamma^0(e_q - g_{\omega q}\omega - \tau_3 g_{\rho q}\rho) - \vec{\gamma}\cdot \vec{p} & & \nonumber \\
 - (m_q - g_{\sigma q}\sigma) -U(r)\bigg]q(\vec{r}) &  =  & 0\ ,
\end{eqnarray}
where $g_{\sigma q}$, $g_{\omega q}$, and $g_{\rho q}$ are the quark coupling constants with $\sigma$, $\omega$, and $\rho$ mesons, with $q$ representing up and down quarks. $m_q$ and $e_q$ represent the mass and energy of quarks, respectively. The confining potential $U(r)$ has a harmonic oscillator form:
\begin{eqnarray}
U(r) = \frac{1}{2}(1+\gamma^0)(ar^2+V_0) \ ,
\end{eqnarray}
where the parameters $a$ and $V_0$ are constants which can be determined by
reproducing reasonable nucleon mass $M_{N}$ and radius $R$ in
free space. With two new defined quantities $m_q^\ast$, $e_q^\ast$\ , the equation can be written as a more compact form:
\begin{eqnarray}
\bigg(-i\bm{\alpha}\cdot \bm{\nabla} + U(r) + \beta m_q^\ast\bigg)q(\vec{r}) = e_q^\ast \ ,
\end{eqnarray}
where $$m_q^\ast = m_q-g_{\sigma q}\sigma \ ,\ \ \ e_q^\ast = e_q-g_{\omega q}\omega-\tau_3 g_{\rho q}\rho \ .$$ This equation can be solved exactly and its ground state solution for energy is
\begin{eqnarray}
(\mathop{e'_q-m'_q})\sqrt{\frac{\lambda_q}{a}} = 3 \ ,
\end{eqnarray}
where $$\lambda_q = e_q^\ast+m_q^\ast\ ,\ \mathop{e'_q} = e_q^\ast-V_0/2\ ,\ \mathop{m'_q} = m_q^\ast+V_0/2\ ,$$
and the wavefunction of quarks is given by
\begin{eqnarray}
\Psi(r,\theta,\phi) = \frac{1}{r} \left(
\begin{array}{c}
F(r)Y_{1/2 m}^0(\theta,\phi) \\
iG(r)Y_{1/2 m}^1(\theta,\phi)
\end{array}
\right),
\end{eqnarray}
where
\begin{eqnarray}
F(r) = \mathcal{N}\left(\frac{r}{r_{0}}\right)\exp(-r^2/2r_0^2)\ , \\ G(r) = -\frac{\mathcal{N}}{\lambda_qr_0}\left(\frac{r}{r_{0}}\right)^2\exp(-r^2/2r_0^2)\ , \nonumber
\end{eqnarray}
\begin{eqnarray}
r_0 = (a\lambda_q)^{-1/4}\ ,\ \ \ \mathcal{N}^2 = \frac{8\lambda_q}{\sqrt{\pi}r_0}\frac{1}{3\mathop{e'_q}+\mathop{m'_q}}\ .
\end{eqnarray}

With the phenomenological potential, the zeroth-order energy of the nucleon core $E_N^0 = \sum_q e_q$ is obtained. But some other effects that are not taken into account inside the nucleon also play important roles. In this work, the first-order contribution of $c.m.$ correction, pionic correction, and gluonic correction are taken into account for the nucleon core energy.

For the $c.m.$ correction, the energy contribution can be written as
\begin{eqnarray}
\epsilon_{c.m.} = \frac{\mathop{77e'_q + 31m'_q}}{3(\mathop{3e'_q + m'_q})^2r_0^2}\ .
\end{eqnarray}

For pionic correction,
\begin{eqnarray}
\delta M_N^\pi = -\frac{171}{25}I_\pi f^2_{NN\pi}\ ,
\end{eqnarray}
where
\begin{eqnarray}
I_\pi = \frac{1}{\pi m_\pi^2}\int_0^{\infty}dk\frac{k^4u^2(k)}{k^2+m_\pi^2}\ , \nonumber
\end{eqnarray}
\begin{eqnarray}
u(k) = \left[1-\frac{3}{2}\frac{k^2}{\lambda_q(\mathop{5e'_q + 7m'_q})}\right]\exp(-\frac{1}{4}r_0^2k^2)\ ,
\end{eqnarray}
and
\begin{eqnarray}
f_{NN\pi} = \frac{25e'_q + 35m'_q}{27e'_q + 9m'_q} \frac{m_\pi}{4\sqrt{\pi}f_\pi}\ . \nonumber
\end{eqnarray}
The constants $m_\pi = 140$ MeV and $f_\pi = 93$ MeV are the pion mass and the phenomenological pion decay constant, respectively.

For gluonic correction,
\begin{eqnarray}
(\Delta E_N)_g = -\alpha_c\left(\frac{256}{3\sqrt{\pi}}\frac{1}{R_{uu}^3}\frac{1}{(\mathop{3e'_q + m'_q})^2}\right)\ ,
\end{eqnarray}
where
\begin{eqnarray}
R_{uu}^2 = \frac{6}{\mathop{e'^2_q - m'^2_q}}\ ,
\end{eqnarray}
and $\alpha_c = 0.58$ is a constant.

With these corrections on energy, the nucleon mass can be determined:
\begin{eqnarray}
M^\ast_N = E^{0}_N-\epsilon_{c.m.}+\delta M_N^\pi+(\Delta E_N)_g\ .
\end{eqnarray}

The nucleon radius in QMF model is written as
\begin{eqnarray}
\langle r_N^2\rangle  =  \frac{\mathop{11e'_q + m'_q}}{\mathop{(3e'_q + m'_q)(e'^2_q-m'^2_q)}}\ .
\end{eqnarray}

Two parameters in the confining potential of Eq.~(2) are then determined to be $a = 0.534296\ \rm{fm}^{-3}$ and $V_0 = -62.257187$ MeV, by reproducing the nucleon mass $M_N = 939$ MeV and radius $r_N = 0.87$ fm.

\subsection{QMF with bag}

In the QMFB model, an additional bag is introduced in describing the nucleon system. It takes into account both the potential and bag to describe the interaction of constituent quarks (with masses of $300$ MeV) in nucleon. The potential again has a harmonic oscillator form and the bag excludes the probability of finding a quark outside a certain region. The Dirac equation for quarks shares the same form with Eq.~(1), but the potential $U(r)$ is different,
\begin{eqnarray}
U(r) = \left\{
\begin{array}{ll}
\frac{1}{2}(1+\gamma^0)(ar^2-aR^2) & r<R \ , \\
\infty & r>R \ ,
\end{array}
\right.
\end{eqnarray}
where $R$ is the radius of the bag (treated as the radius of nucleon) and $a=0.534296\ \rm{fm}^{-3}$, same as in the model without bag. When $a\rightarrow 0$, the QMFB model goes back to QMC.

The solution for the Dirac equation of quarks inside the bag ($r<R$) can be written as:
\begin{eqnarray}
\Psi(r,\theta,\phi) = \begin{pmatrix}
\frac{F(r)}{r}Y_{jm}^l(\theta,\phi) \\ i\frac{G(r)}{r}Y_{jm}^{\tilde{l}}(\theta,\phi)
\end{pmatrix}
 \ .
\end{eqnarray}

For ground state, $n = 1$, $l = 0$, $\tilde{l} = 1$. Then the radial equation for $F(r)$ is
\begin{eqnarray}
\frac{d^2F(r)}{dr^2} - \frac{l(l+1)}{r^2}F(r)  =  \lambda_q[m_q^\ast - e^\ast_q + U(r)]F(r) \ .
\end{eqnarray}
We rewrite this equation with $H(r) = F(r)/r$,
\begin{eqnarray}
\frac{d^2H(r)}{dr^2} + \frac{2}{r}\frac{dH(r)}{dr} + \biggl[(e^\ast_q + m_q^\ast)(e^\ast_q - m_q^\ast + aR^2) \nonumber \\
- a(e^\ast_q + m_q^\ast)r^2 - \frac{l(l+1)}{r^2} \biggr]H(r)  =  0\ .
\end{eqnarray}
When $r\rightarrow 0$, the equation becomes
\begin{eqnarray}
\frac{d^2H(r)}{dr^2} + \frac{2}{r}\frac{dH(r)}{dr} - \frac{l(l+1)}{r^2}H(r)  =  0 \ ,
\end{eqnarray}
and $$H \propto r^l\ \ \ \text{or}\ \ \ H \propto r^{-(l+1)} \ .$$
The second term gives an infinite value at $r = 0$, so we neglect it.
When $r\rightarrow \infty$, the equation becomes
\begin{eqnarray}
\frac{d^2H(r)}{dr^2} - a(e_q^\ast + m_q^\ast)r^2H(r)  =  0 \ ,
\end{eqnarray}
and $$H \propto \exp(\pm\alpha r^2/2)\ ,$$
where $\alpha  =  -\sqrt{a(e_q^\ast + m_q^\ast)}$\ . We neglect $\exp(-\alpha r^2/2)$ because it approaches to infinity when $r\rightarrow \infty$.
Combining these asymptotic behaviors, we have a form of $H(r)$ as
\begin{eqnarray}
H(r)  =  Nr^l \exp(\alpha r^2/2) u(r) \ .
\end{eqnarray}
When substituting this formula into Eq.~(18), we obtain
\begin{eqnarray}
\frac{d^2u(r)}{dr^2} + \frac{2}{r}(l+1+\alpha r^2)\frac{du(r)}{dr} & & \nonumber \\
+ [\kappa + (2l+3)\alpha]u(r) &  =  & 0 \ ,
\end{eqnarray}
where $\kappa  =  (e_q^\ast + m_q^\ast)(e_q^\ast - m_q^\ast + aR^2)$.
Subsequently, we use $\xi$ to represent $-\alpha r^2$ and rewrite Eq.~(22) as
\begin{eqnarray}
\xi\frac{d^2u(r)}{d\xi^2} + (l+3/2-\xi)\frac{du(r)}{d\xi} & & \nonumber \\
+ \biggl[-\frac{\kappa}{4\alpha} - \frac{(l+3/2)}{2} \biggr]u(r) &  =  & 0 \ ,
\end{eqnarray}
where $\lambda  =  \frac{\kappa}{4\alpha} + \frac{l+3/2}{2}$ and $\beta  =  l + 3/2$. This is a confluent hypergeometric equation and its solution can be written as $\ _1F_1(\lambda,\beta,\alpha r^2)$. The solution $\ _1F_1$ is called confluent hypergeometric function and it is composed of an infinite series, $$\ _1F_1(\lambda, \beta, \alpha r^2)  =  \sum_{n = 0}\frac{\lambda^{(n)} (\alpha r^2)^n}{\beta^{(n)} n!},$$ where $\lambda^{(n)} = \lambda(\lambda+1)(\lambda+2)\cdots (\lambda+n-1)$. So the solution of ground state for Eq.~(17) is
\begin{eqnarray}
F(r)  =  Nr\exp(\alpha r^2/2)\ _1F_1(\eta,\frac{3}{2},-\alpha r^2) \ ,
\end{eqnarray}
where $\eta = \kappa/(4\alpha) + 3/4$.

$G(r)$ in Eq.~(16) is
\begin{eqnarray}
G(r)  =  -\frac{aN}{\alpha}r^2\exp(\alpha r^2/2) \bigg[\ _1F_1(\eta,\frac{3}{2},-\alpha r^2) \nonumber \\
+\frac{4\eta-6}{3}\ _1F_1(\eta,\frac{5}{2},-\alpha r^2)\bigg] \ ,
\end{eqnarray}
where $N$ is the normalizing constant.

The fact of prohibiting quarks flux outside the bag demands a boundary condition $|F(R)|^2  =  |G(R)|^2$, or
\begin{eqnarray}
\alpha^2\bigg|\ _1F_1(\eta,\frac{3}{2},-\alpha r^2)\bigg|^2  =  a^2 R^2\bigg[\ _1F_1(\eta,\frac{3}{2},-\alpha r^2) \nonumber \\
+ \frac{2(\eta-\beta)}{\beta}\ _1F_1(\eta,\frac{5}{2},-\alpha r^2)\bigg]^2 \ .
\end{eqnarray}
This equation can be used to determine the kinetic energy of quarks $e_q^\ast$.

The total energy or nucleon mass can be written as:
\begin{eqnarray}
M_N  =  3e_q^\ast - \frac{Z_0}{R} + \frac{4}{3}\pi R^3 B \ ,
\end{eqnarray}
where $Z_0$ is the zero-point energy parameter and $B$ is the bag constant. These two parameters will be determined by the mass and radius of nucleon in free space. within the numerical procedure, we first give the model parameters: $m_q$, $a$, $Z_0$, $B$, then calculate the nucleon mass using Eq.~(27) by varying the radius $R$. After getting the nucleon mass as a function of $R$, we determine the resulting mass $M_{N}$ and radius $R$ by the stable condition,
\begin{eqnarray}
\frac{d M_{N} (R)}{dR} =0.
\end{eqnarray}
Finally $Z_0 = 4.5445260$ and $B = 64.5136659\ \rm{MeV/fm^3}$ are obtained by fitting the mass and radius of free nucleon.

\section{Nuclear Matter and Symmetric energy}

To carry out the study of many-body problems in QMF model, the meson-exchange scenario is employed to describe the nuclear interaction. The meson-coupling Lagrangian is given as:
\begin{eqnarray}
\mathcal{L}&  =  & \overline{\Psi}_N\left(i\gamma_\mu \partial^\mu - M_N^\ast - g_{\omega N}\omega\gamma^0 - g_{\rho N}\rho\tau_{3}\gamma^0\right)\Psi_N \nonumber \\
& & -\frac{1}{2}(\nabla\sigma)^2 - \frac{1}{2}m_\sigma^2 \sigma^2 - \frac{1}{3}g_2\sigma^3 - \frac{1}{4}g_3\sigma^4 \nonumber \\
& & + \frac{1}{2}(\nabla\rho)^2 + \frac{1}{2}m_\rho^2\rho^2 + \frac{1}{2}(\nabla\omega)^2 + \frac{1}{2}m_\omega^2\omega^2 \nonumber \\
& & + \frac{1}{2}g_{\rho N}^2\rho^2 \Lambda_v g_{\omega N}^2\omega^2 \ ,
\end{eqnarray}
where $m_\sigma$, $m_\omega$, and $m_\rho$ denote the mass of sigma, omega, and rho mesons, respectively. Their values, $m_\sigma  =  510$, $m_\omega = 783$, and $m_\rho = 770$ MeV, are fixed in this work. $g_{\omega N}$ and $g_{\rho N}$ are the coupling constants of the omega and rho mesons with nucleons. From the quark counting rule, we obtain $g_{\omega N} = 3g_{\omega q}$ and $g_{\rho N} = g_{\rho q}$. The effective nucleon mass $M_N^\ast$ here is a function of sigma field and depends on the parameter $g_\sigma^q$, which is determined by the previous step. The last term of the Lagrangian is the coupling term of the omega meson and rho meson. It is added to achieve a reasonable value of the symmetric energy slope at saturation  density. The nonlinear terms of sigma meson are introduced here, while that of omega meson is not. The reason, as explained in Ref.~\cite{zhu18}, is to fulfill the observed 2-solar-mass constraint~\citep{0348+0432,1614-2230-1,1614-2230-2} with empirical saturation properties.

The QMF Lagrangian has six parameters, which are determined by reproducing emperical properties of nuclear matter at saturation density. Six quantities that measured by terrestrial experiments are chosen: the saturation density $\rho_0$; the binding energy $E/A$; the symmetric energy $E_{\rm sym}$; incompressibility $K$; the symmetry energy slope $L$; and the effective mass $M_N^\ast$. The experimental values for these quantities are collected in Table I. In particular, we use the most preferred values for $(K,~J,~L)$ as recently suggested in Refs.~\cite{baoan,steiner}, namely $K  =  240 \pm 20$ MeV, $J  =  31.6 \pm 2.66$ MeV, and $L  =  58.9 \pm 16$ MeV. We then fit 12 parameter sets, 6 for QMF with bag and 6 for QMF without bag, for the study of nuclear matter and neutron stars.

The obtained ($g_{\sigma q},~g_{\omega q},~g_{\rho q},~g_2,~g_3,~\Lambda_v$) values for each parameter set are shown in Table II. One can see that $\omega$, $\rho$ coupling constants stay unchanged after the bag is included in the model, which demonstrates that the bag merely affects the scalar meson. It is reasonable because the bag is introduced in describing properties of nucleons, which is merely correlated with the $\sigma$ coupling.

\begin{table}\label{tab1}
\caption{Saturation properties used in this study for the fitting of new sets of meson coupling parameters: The saturation density $\rho_0$ (in fm$^{-3}$) and the corresponding values at saturation point for the binding energy $E/A$ (in MeV), the incompressibility $K$ (in MeV), the symmetry energy $E_{\rm sym}$ (in MeV), the symmetry energy slope $L$ (in MeV) and the ratio between the effective mass and free nucleon mass $M_N^\ast/M_N$.}
\begin{center}
\begin{tabular}{ccccccc} \hline
$\rho_0$ & $E/A$ & $E_{\rm sym}$ & $K$ & $L$ & $M_N^\ast/M_N$ \\
$[{\rm fm}^{-3}$] & [MeV] & [MeV] & [MeV] & [MeV] & / \\ \hline
$0.16$ & $-16.0$ & $31.0$ & $240.0$/$260.0$ & $40.0$/$60.0$/$80.0$ & $0.79$ \\ \hline
\end{tabular}
\end{center}
\end{table}

\begin{table}\label{tab2}
\caption{Newly fitted meson coupling parameters by using Table I as input, for both QMF models with and without bag.}
\begin{center}
\begin{tabular}{c|c|c|c|c|c|c|c} \hline \hline
 \multicolumn{8}{c}{QMF with bag} \\ \hline
$L$ & $K$ & $g_{\sigma q}$ & $g_{\omega q}$ & $g_{\rho q}$ & $g_2$ & $g_3$ & $\Lambda_v$ \\ \hline
$40$ & \multirow{3}{*}{240} & \multirow{3}{*}{$4.0833$} & \multirow{3}{*}{$2.7445$} & $5.3700$ & \multirow{3}{*}{$-14.8681$} & \multirow{3}{*}{$-31.1260$} & $0.9221$ \\ 
$60$ & & & & $4.5831$ & & & $0.5072$ \\ 
$80$ & & & & $4.0643$ & & & $0.09232$ \\ \hline
$40$ & \multirow{3}{*}{260} & \multirow{3}{*}{$4.0340$} & \multirow{3}{*}{$2.7445$} & $5.3638$ & \multirow{3}{*}{$-10.7966$} & \multirow{3}{*}{$-12.2786$} & $0.9195$ \\
$60$ & & & & $4.5793$ & & & $0.5046$ \\
$80$ & & & & $4.0616$ & & & $0.08975$ \\ \hline \hline

 \multicolumn{8}{c}{QMF without bag} \\ \hline
$L$ & $K$ & $g_{\sigma q}$ & $g_{\omega q}$ & $g_{\rho q}$ & $g_2$ & $g_3$ & $\Lambda_v$ \\ \hline
$40$ & \multirow{3}{*}{240} & \multirow{3}{*}{$3.7488$} & \multirow{3}{*}{$2.7445$} & $5.3700$ & \multirow{3}{*}{$-17.0884$} & \multirow{3}{*}{$-60.9866$} & $0.9221$ \\
$60$ & & & & $4.5831$ & & & $0.5072$ \\
$80$ & & & & $4.0643$ & & & $0.09232$ \\ \hline
$40$ & \multirow{3}{*}{260} & \multirow{3}{*}{$3.7058$} & \multirow{3}{*}{$2.7445$} & $5.3638$ & \multirow{3}{*}{$-13.1520$} & \multirow{3}{*}{$-41.8332$} & $0.9195$ \\ 
$60$ & & & & $4.5793$ & & & $0.5046$ \\ 
$80$ & & & & $4.0616$ & & & $0.08975$ \\ \hline
\end{tabular}
\end{center}
\end{table}

By variation of the Lagrangian, the equations for the measons in mean-field approximation are obtained:
\begin{eqnarray}
m_\sigma^2\sigma + g_2\sigma^2 + g_3\sigma^3 &  =  & -\frac{\partial M_N^\ast}{\partial \sigma}\rho_S\ , \\
m_\omega^{\ast2}\omega &  =  & g_{\omega N}\rho_N \ , \\
m_\rho^{\ast2}\rho &  =  & g_{\rho N}\rho_3\ ,
\end{eqnarray}
where
\begin{eqnarray}
\rho_S &  =  &  \frac{1}{\pi^2}\sum_{i = n,p}\int_0^{p_F^i}dpp_i^2\frac{M^\ast_N}{\sqrt{M^{\ast2}_N+p_i^2}} \nonumber \\
 &  =  &  \frac{M_N^\ast}{2\pi^2} \left(p_F^iE_F^i - M_N^{\ast2}\ln\bigg|\frac{p_F^i+E_F^i}{M_N^\ast}\bigg|\right)\ ,\nonumber
\end{eqnarray}
\begin{eqnarray}
E_F^i = \sqrt{M_N^{\ast2}+(p_F^i)^2} \nonumber\ , \\
m_\omega^{\ast2} = m_\omega^2+\Lambda_v g_{\omega N}^2g_{\rho N}^2\rho^2\ , \nonumber \\
m_\rho^{\ast2} = m_\rho^2+ \Lambda_v g_{\rho N}^2g_{\omega N}^2\omega^2\ . \nonumber
\end{eqnarray}
$p_F^n~(p_F^p$) is the Fermi momentum for neutrons (protons), $\rho_N = \rho_p + \rho_n$ and $\rho_3 = \rho_p - \rho_n$ that equals $0$ in symmetric nuclear matter.

The values of meson fields will be obtained by solving Eqs.~(30)$-$(31). The energy density and pressure can be generated by Legendre transformation from the Lagrangian:
\begin{eqnarray}
\mathcal{E} &  =  & \frac{1}{\pi^2}\sum_{i = n,p}\int_0^{p_F^i}\sqrt{p^2+M_N^{\ast2}}p^2dp+ g_{\omega N}\omega\rho_B  \nonumber \\
& &  + g_{\rho N}\rho\rho_3 + \frac{1}{2}m_\sigma^2\sigma^2 + \frac{1}{3}g_2\sigma^3 + \frac{1}{4}g_3\sigma^4
 - \frac{1}{2}m_\omega^2\omega^2 \nonumber \\
 & & - \frac{1}{2}m_\rho^2\rho^2 - \frac{1}{2}\Lambda_vg_{\rho N}^2g_{\omega N}^2\rho^2\omega^2 \ , \\
P &  =  & \frac{1}{3\pi^2}\sum_{i = n,p}\int_0^{p_F^i}\frac{p^4}{\sqrt{p^2+M_N^{\ast2}}}dp - \frac{1}{2}m_\sigma^2\sigma^2 \nonumber \\
& & - \frac{1}{3}g_2\sigma^3 - \frac{1}{4}g_3\sigma^4 + \frac{1}{2}m_\omega^2\omega^2 + \frac{1}{2}m_\rho^2\rho^2 \nonumber \\
& & + \frac{1}{2}\Lambda_vg_{\rho N}^2g_{\omega N}^2\rho^2\omega^2 \ .
\end{eqnarray}
Then the properties of nuclear matter can be determined.

We then write down the expressions of other quantities in Table I. The symmetry energy is determined by
\begin{eqnarray}
E_{\rm{sym}} = \frac{1}{2}\frac{\partial^2E(\rho_N,\beta)}{\partial \beta^2}\bigg|_{\beta = 0} = \frac{p_F^2}{6E_F}+\frac{g_{\rho N}^2}{2m_\rho^{\ast2}}\rho_N \ ,
\end{eqnarray}
where $\beta = \frac{\rho_n-\rho_p}{\rho_N}$ is called neutron-excess parameter, $E(\rho_N,\beta)$ is the binding energy, $p_F = p_F^n = p_F^p$ and $E_F = E_F^n = E_F^p.$ The incompressibility $K$ and symmetry energy slope are determined by
\begin{eqnarray}
K_0 &  =  & 9\frac{dP}{d\rho_B}\bigg|_{\beta = 0,\rho_N = \rho_0}\nonumber \\
&  =  & \frac{3p_F^2}{E_F}+\frac{3M_N^\ast p_F}{E_F}\frac{dM_N^\ast}{dp_F} + \frac{9g_{\omega N}^2}{m_\omega^2}\rho_0 \ .
\end{eqnarray}

\begin{eqnarray}
L = 3\rho_0\frac{\partial E_{\rm{sym}}(\rho_N)}{\partial \rho_N}\bigg|_{\rho_N = \rho_0}\ .
\end{eqnarray}
Note here that the symmetry energy slope $L$ is solely determined by $g_{\rho q}$ and $\Lambda_v$. This dependence of $L$ is obviously in Eq.~(35) and Eq.~(37).

The effective nucleon mass and radius as functions of baryon density for symmetric nuclear matter are displayed in Fig.~1. We compared the results with and without bags for different values of $K$. One may notice that the effects of $K$ and bag on effective mass are weak at density below $\sim2\rho_0$, which is largely because the effective mass is fixed at saturation density for all parameter sets. The bag effects become evident at density higher than $\sim2\rho_0$ and clearly decrease the effective mass. However, the influence of $K$ on the effective mass is weaker and negligible even at high density in the case without bag. For the radius of the nucleon, the two models, with and without bag, show different tendencies. The radius keeps increasing with density in the model without bag, while in the model with bag, the radius increases first, then after a peak value ($\sim0.883$ fm) it begins to decrease. From this and also our previous work~\cite{rN}, we can argue that in the QMFB model for the contribution from quark confinement, the potential may be important at low density, while the bag becomes dominant when the density is high.

\begin{figure}
\vspace{0.3cm}
{\centering
\resizebox*{0.48\textwidth}{0.3\textheight}
{\includegraphics{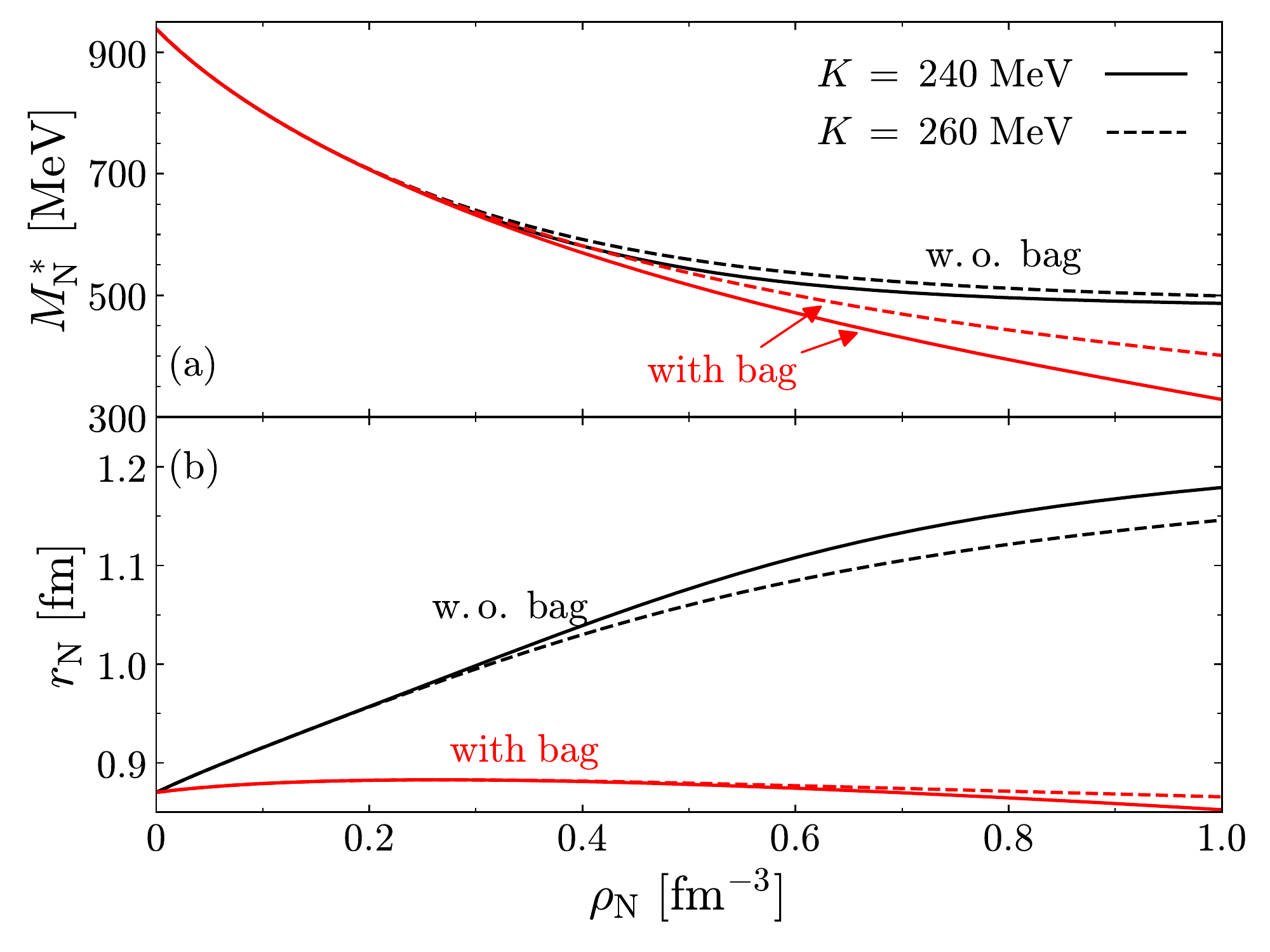}}
\par}
\caption{\small(Color online) Effective nucleon mass (upper panel) and radius (lower panel) as functions of baryon density for nuclear matter. The model with bag (red) and without bag (black), $K  =  240$ MeV (solid) and $K  =  260$ MeV (dashed) are included for comparison.}\label{fig1}
\end{figure}

The binding energy per baryon and pressure as functions of density for nuclear matter are displayed in Fig.~2. The pink region denotes the constraint from heavy-ion collision (HIC) experiment~\cite{HICsci}. Both the models with bag and without bag can satisfy the HIC constraint. A large $K$ apparently increases the binding and pressure because $K$ is defined as the derivative of pressure of nuclear matter at saturation density. However its influence is much weaker than that of the bag and is suppressed in the model without bag. The inclusion of bag clearly decreases the binding energy and pressure since it enhances the attractive $\sigma-N$ coupling, as can be seen from the upper panel of Fig.~1.

\begin{figure}
\vspace{0.3cm}
{\centering
\resizebox*{0.48\textwidth}{0.3\textheight}
{\includegraphics{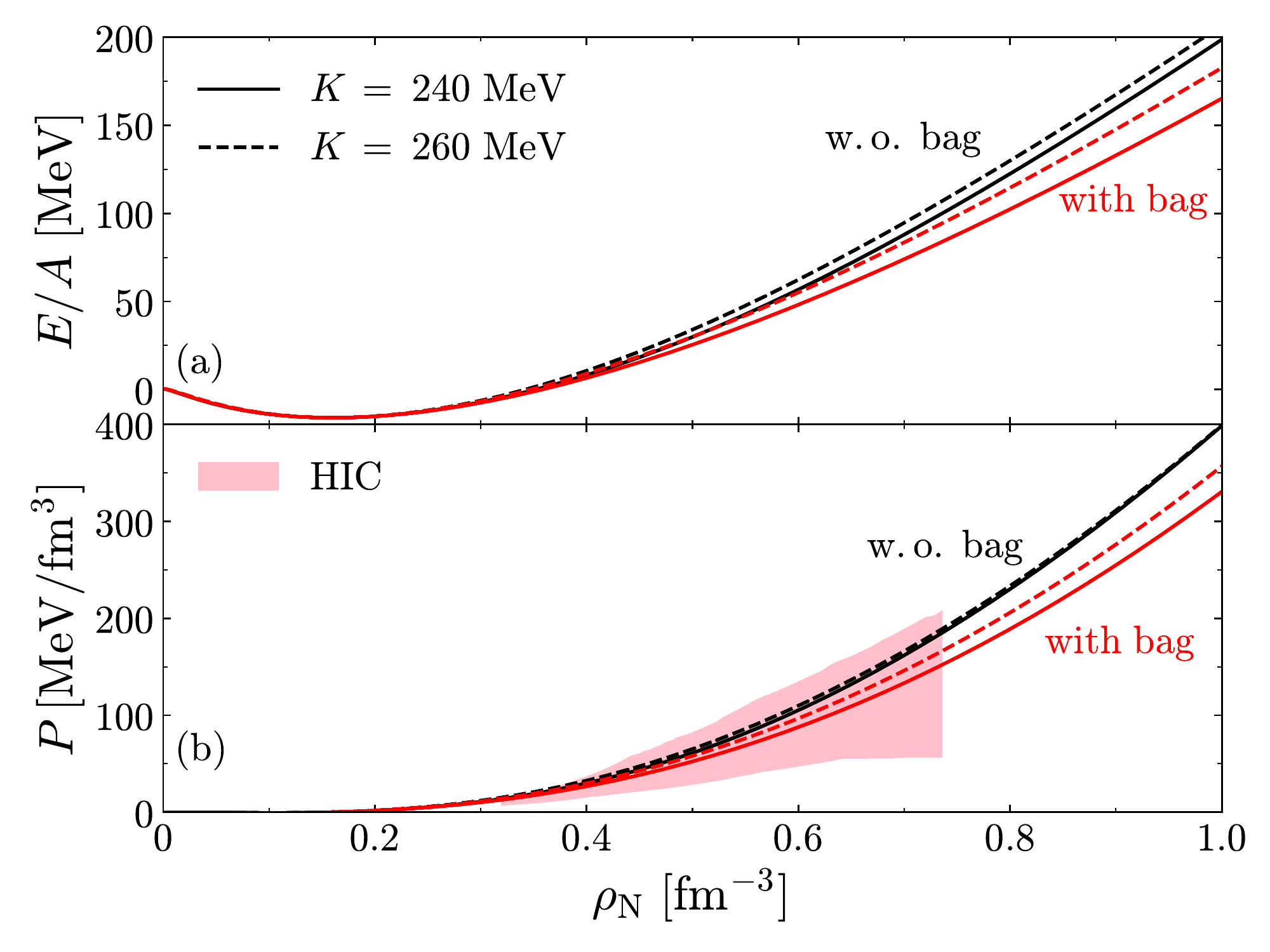}}}
\caption{\small(Color online) Density dependence of binding energy per nucleon (upper panel) and pressure (lower panel) for nuclear matter. The denotation for models and $K$ is the same with previous figure.}\label{fig2}
\end{figure}

Before the discussions on neutron stars, we present in Fig.~3 the symmetry energy as a function of density. Results for $L  =  40$, $60$, and $80$ MeV for both QMF models are presented, with fixed $K  = 240$ MeV for all sets. Various laboratory constraints~\cite{IAS,alphaD,HICsym} are also shown. One can see that the symmetry energy at subsaturation density is weakly affected by the bag, so the present experimental data can effectively constrain the value of $L$ in the present model. Among all the cases, we notice that the set with $L  =  40$ MeV lies inside all experimental boundaries. The bag effects becomes considerable at high density. To be opposite with the pressure case as shown in Fig.~2, the symmetry energy is stiffened when the bag is included in the model. This effect from the bag can be clearly observed in Eq.~(35). The second term is identical for both cases since the coupling constants $g_{\rho q}$ and $\Lambda_v$ are independent on the models, while the first term are affected by the effective mass of nucleon. A less effective mass obtained in QMFB results in less nucleon energy and consequently enhance the symmetry energy.

\begin{figure}
\vspace{0.3cm}
{\centering
\resizebox*{0.48\textwidth}{0.3\textheight}
{\includegraphics{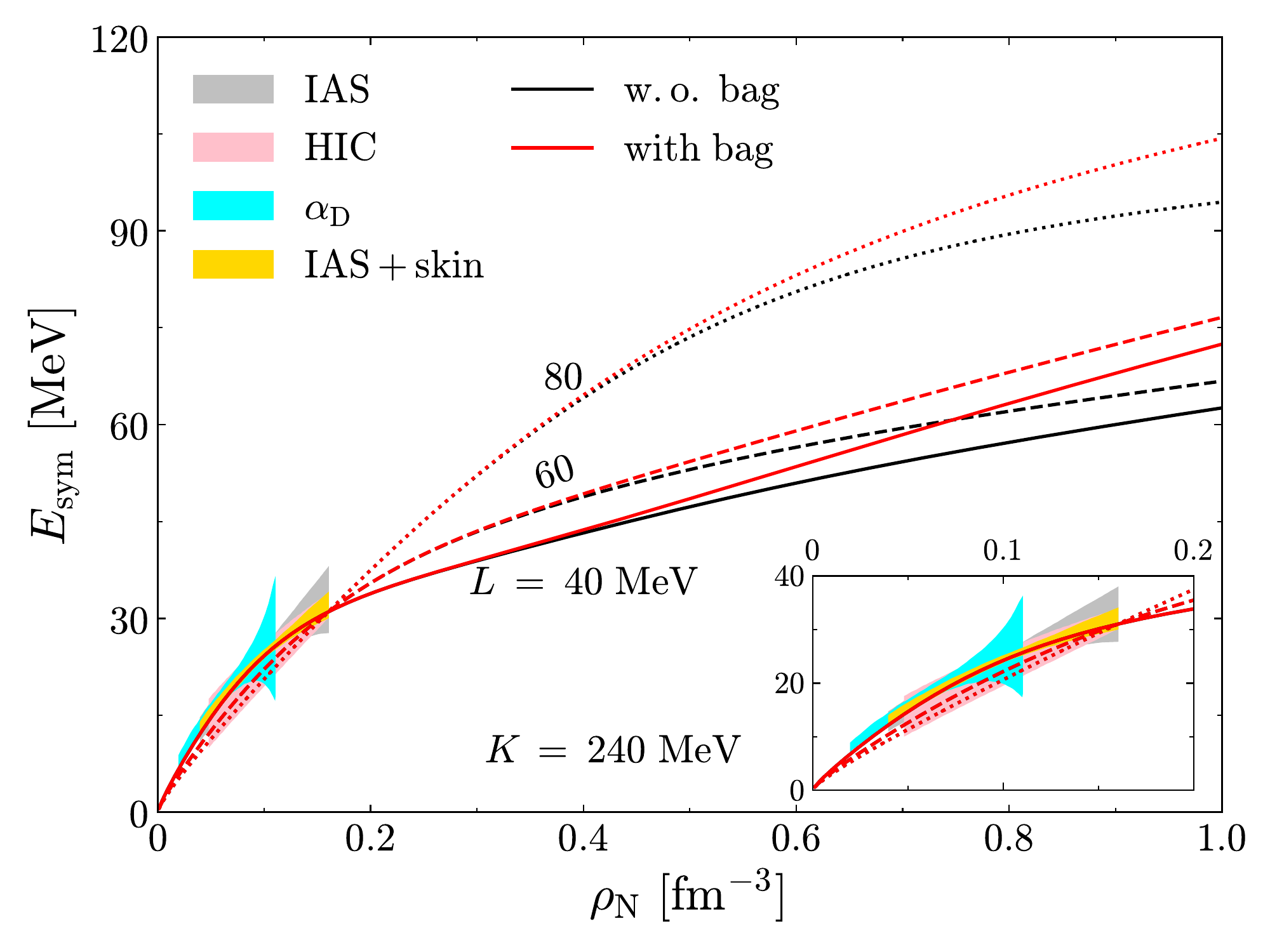}}
\par}
\caption{\small(Color online) Symmetry energy as a function of density, with $L  =  40$ (solid), $60$ (dashed) and $80$ MeV (dotted) for both QMF models with bag (red) and without bag (black). The incompressibility $K$ is fixed at $240$ MeV for all sets. The colorful shadow regions denote the constraints from IAS (silver) and from IAS+skin (gold)~\cite{IAS}, from the $\alpha_{\rm D}$ (cyan)~\cite{alphaD} and from HIC (pink)~\cite{HICsym}, respectively.}\label{fig3}
\end{figure}

\section{Neutron Star}

in order to study of neutron star matter, the contribution from leptons has to be included,
\begin{eqnarray}
\mathcal{L}_{\rm lep}&  =  & \sum_{l = e,\mu}\overline{\Psi}_l\left(i\gamma_\mu \partial^\mu - m_l \right)\Psi_l \ .
\end{eqnarray}
The contributions from leptons for energy density and pressure are
\begin{eqnarray}
\mathcal{E}_{\rm lep} &  =  & \frac{1}{\pi^2}\sum_{l = e,\mu}\int_0^{p_F^l}\sqrt{p^2+m_l^2}p^2dp \ ,
\end{eqnarray}
\begin{eqnarray}
P_{\rm lep} &  =  & \frac{1}{3\pi^2}\sum_{l = e,\mu}\int_0^{p_F^l}\frac{p^4}{\sqrt{p^2+m_l^2}}dp \ ,
\end{eqnarray}
respectively. Then the conditions of $\beta$-equilibrium and charge neutrality have to be considered,
\begin{eqnarray}
\mu_p+\mu_e = \mu_n\ ,\ \ \ \mu_e = \mu_\mu\ ,
\end{eqnarray}
\begin{eqnarray}
\rho_p-\rho_e-\rho_\mu = 0 \ ,
\end{eqnarray}
where $\mu_p,\ \mu_n,\ \mu_e$, and $\mu_\mu$ denote the chemical potential of protons, neutrons, electrons, and muons, respectively. The equation of state (EoS) of the neutron star matter can then be computed through Eqs.~(33)$-$(34) and Eqs.~(39)$-$(40) after solving the Eqs.~(30)$-$(32) and Eqs.~(41)$-$(42). After that, the stable configurations of a neutron star can be obtained by solving the hydrostatic equilibrium Tolman-Oppenheimer-Volkoff (TOV) equations,
\begin{eqnarray}
\frac{dP(r)}{dr}& = &-\frac{Gm(r)\varepsilon(r)}{r^{2}}\frac{\Big[1+\frac{P(r)}{\varepsilon(r)}\Big]\Big[1+\frac{4\pi r^{3}P(r)}{m(r)}\Big]}
 {1-\frac{2Gm(r)}{r}} \ ,\\
\frac{dm(r)}{dr}& = &4\pi r^{2}\varepsilon(r) \ .
\end{eqnarray}

\begin{figure}
\vspace{0.3cm}
{\centering
\resizebox*{0.48\textwidth}{0.3\textheight}
{\includegraphics{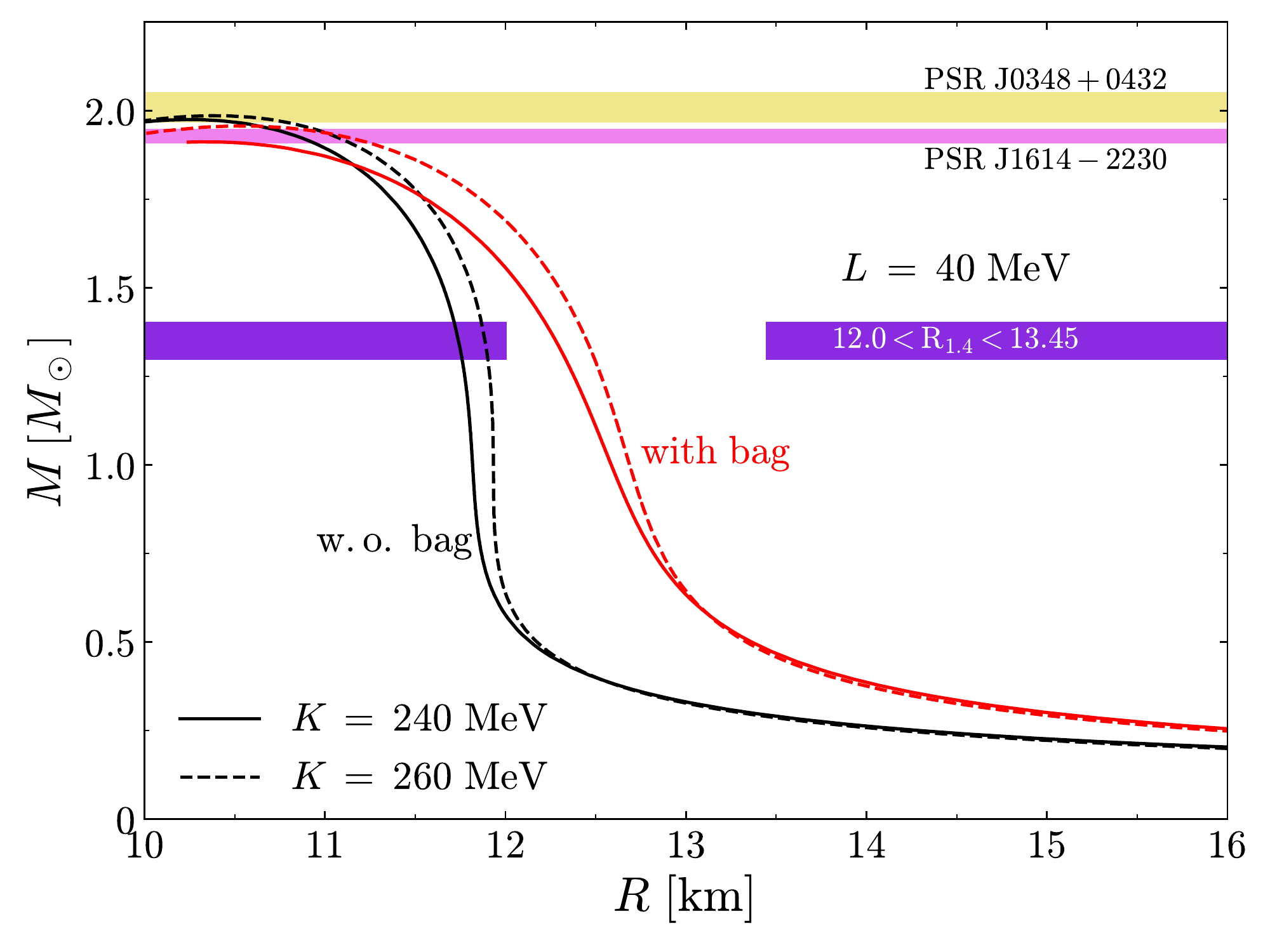}}
\par}
\caption{\small(Color online) Mass-radius relations of neutron stars for QMF models with bag (red) and without bag (black). The sets with different value of $K$ are also included for comparison, while the $L$ is fixed at $40$ MeV. The colorful shadow regions represent the measured mass of two massive neutron stars $\rm{PSR\ J}0348+0432$~\cite{0348+0432} and $\rm{PSR\ J}1614-2230$~\cite{1614-2230-1,1614-2230-2} and the radius constraints for a $1.4M_\odot$ star from GW170817~\cite{Elias}.}\label{fig4}
\end{figure}

\begin{figure}
\vspace{0.3cm}
{\centering
\resizebox*{0.48\textwidth}{0.3\textheight}
{\includegraphics{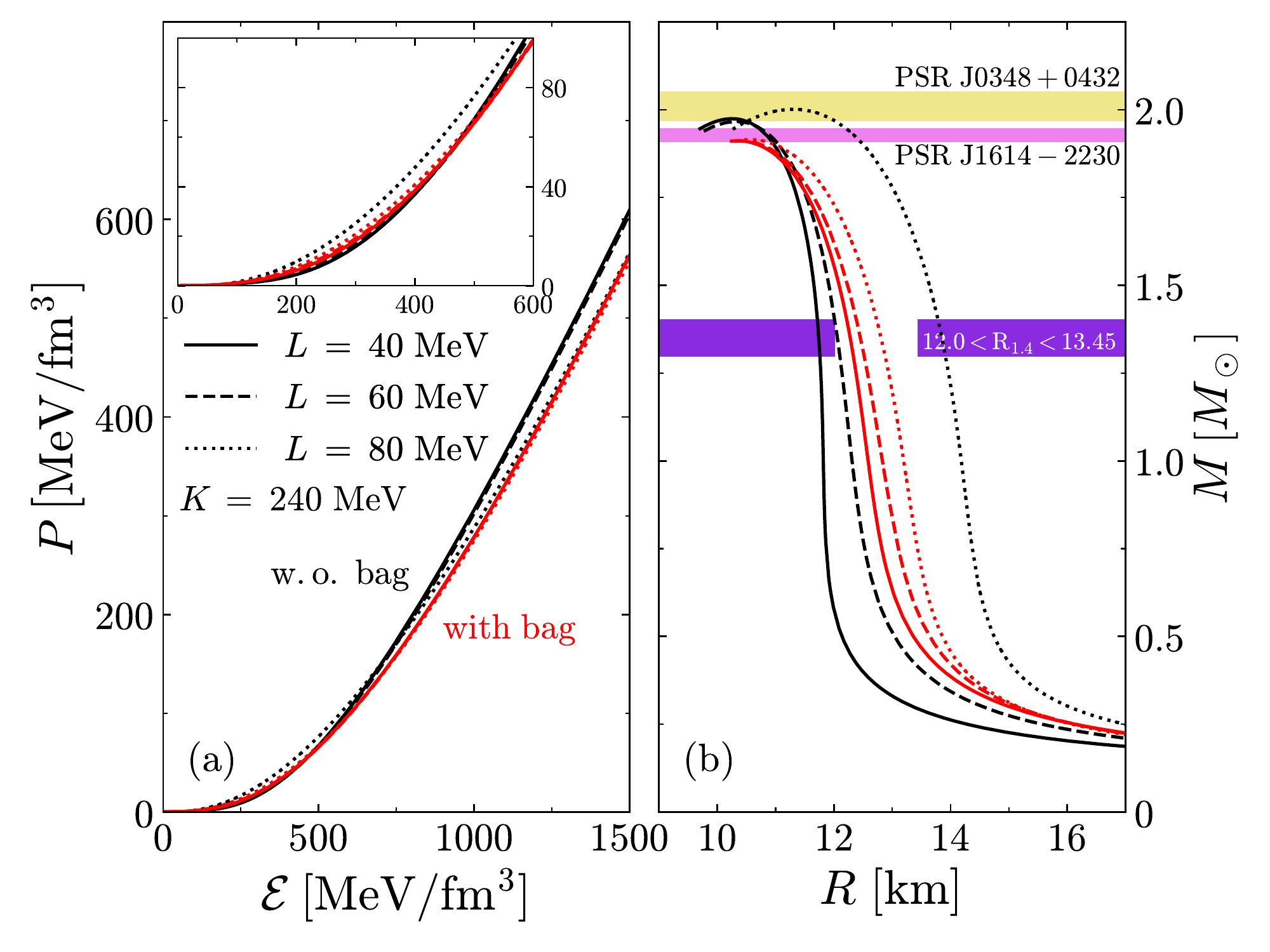}}
\par}
\caption{\small(Color online) EoSs (left panel) and mass-radius relations (right panel) of neutron stars for QMF models with bag (red) and without bag (black), with fixed $K  =  240$ MeV and three values of $L  =  40,~60,~80$ MeV. The radius constraints are also shown as in Fig.~4.}\label{fig5}
\end{figure}

The mass-radius relations of neutron stars for QMF models with and without bag are displayed in Fig.~4, with fixed $L  =  40$ MeV and two values of $K$, namely $240,~260$ MeV. As expected from Fig.~2, $K$ has a much weaker effects than the bag. Particularly, the bag effects are significant on the radius ($\sim 0.7$ km) though it is small on the maximum mass. This is because that the bag will increase the symmetry energy (as seen from Fig.~3) and stiffen the EoSs of neutron stars below $0.53~\rm{fm^{-3}}$ (a density around the center density of a $1.4~M_\odot$ neutron star). We observe that the new model with bag (i.e., QMFB) is more preferred by the recent radius constraints~\cite{Elias,Fattoyev,Annala,Bauswein} from the gravitational wave event of GW170817, as one can see in Fig.~4 where one of the latest results~\cite{Elias} for a $1.4M_\odot$ star is included.

The symmetry energy presented in Fig.~3 shows considerable effects of $L$, which is essential for determining the neutron star radius, i.e.,  well-accepted $R$ vs $L$ dependence~\cite{zhu18,lvsr1,lvsr2,lvsr3}. Therefore, we further present in Fig.~5 the influence of $L$ on the mass-radius relations, together with the corresponding EoSs. In the left panel the inset further displays the pressure in the low density region, where the maximum value of the energy density is close to the central density of a $1.4\ M_\odot$ star. With $L$ increasing from $40$ MeV to $80$ MeV, the stiffness of the EoS increases, as well as the radius of $1.4M_\odot$ star. Also, it is evident that the $R-L$ dependence is much suppressed in the QMFB model. In the model without bag, the radius is around $11.7$ km with $L  =  40$ MeV, and is increased by $14~\%$ to around $13.7$ km with $L  =  80$ MeV. However in the case of model with bag, the radius is merely increased by $3.9~\%$, with $12.2$ km for $L  =  40$ MeV and $12.7$ km for $L  =  80$ MeV. This is the consequent that the EoS curves with bag are more compact comparing to those without bag, as one may notice in the inset.

\section{Summary}

In this work, we have introduced the bag to describe the confinement of quarks in QMF model, the new QMFB model. In the case of the model with bag, the Dirac equation for quarks is solved and the contributions from zero-point energy and bag constant are included. Furthermore, the nuclear matter and neutron star are studied by employing the meson-exchange scenario and the comparisons are done for models with and without bag. Therefore, 12 new parameter sets for describing the nucleon-nucleon interactions are obtained by fitting the empirical properties ($E/A,~E_{\rm sym},~K,~L,~M_N^\ast/M_N$) of nuclear matter at saturation density $\rho_0$.

We have shown that the confinement is mainly demonstrated by the bag after it is included in QMF model. For nuclear matter, the bag mainly decreases the binding energy and increases the symmetry energy. For neutron star, the radius for $1.4M_\odot$ neutron stars is significantly modified by the bag effects with the maximum mass only slightly modified, similar with the well-known $L$ vs $R$ correlation. Nevertheless, the $L$ effect on radius is suppressed after the bag is introduced in the model.

\section{Acknowledgments}
We would like to thank Ping Wang, Huan Chen and Antonio Figura for valuable discussions. This work was supported in part by the National Natural Science Foundation of China (Grants Nos. 11873040, 11775119, and 11675083).


\begin{thebibliography}{}

\bibitem{emc1}{}
P. R. Norton, Rep. Prog. Phys. {\bf 66}, 1253 (2003).

\bibitem{emc2}{}
D. F. Geesaman, K. Saito, and A. W. Thomas, Annu. Rev. Nucl. Part. Sci. {\bf 45}, 337 (1995).

\bibitem{pro1}{}
T. Mart, and A. Sulaksono, Phys. Rev. C {\bf 87}, 025807 (2013).

\bibitem{pro2}{}
Suparti, A. Sulaksono, and T. Mart, Phys. Rev. C {\bf 95}, 045806 (2017).

\bibitem{rN}{}
Z. Y. Zhu, and A. Li, Phys. Rev. C {\bf 97}, 035805 (2018).

\bibitem{qmcgui}{}
P. Guichon, Phys. Lett. B {\bf 200}, 235 (1988).

\bibitem{qmc1}{}
P. Guichon, K. Saito, E. Rodionov, and A. W. Thomas, Nucl. Phys. A {\bf 601}, 349 (1996).

\bibitem{qmc2}{}
T. Miyatsu, T. Katayama, and K. Saito, Phys. Lett. B {\bf 709}, 242 (2012).

\bibitem{qmc3}{}
P. K. Panda, A. M. S. Santos, D. P. Menezes, and C. Provid\^{e}ncia, Phys. Rev. C {\bf 85}, 055802 (2012).

\bibitem{qmc4}{}
P. K. Panda, M. E. Bracco, M. Chiapparini, E. Conte, and G. Krein, Phys. Rev. C {\bf 65}, 065206 (2002).

\bibitem{tokiqmf}{}
H. Toki, U. Meyer, A. Faessler, and R. Brockmann, Phys. Rev. C {\bf 58}, 3749 (1998).

\bibitem{shenqmf}{}
H. Shen, and H. Toki, Phys. Rev. C {\bf 61}, 045205 (2000).

\bibitem{wangp1}{}
P. Wang, Z. Y. Zhang, Y. W. Yu, R. K. Su, and Q. Song, Nucl. Phys. A {\bf 688}, 791 (2001).
\bibitem{wangp2}{}
P. Wang, D. B. Leinweber, A. W. Thomas, and A. G. Williams, Nucl. Phys. A {\bf 744}, 273 (2004).

\bibitem{qmf1}{}
J. N. Hu, A. Li, H. Shen, and H. Toki, Prog. Theor. Exp. Phys. {\bf 2014}, 013D02 (2014).

\bibitem{qmf2}{}
J. N. Hu, A. Li, H. Toki, and W. Zuo, Phys. Rev. C {\bf 89}, 025802 (2014).

\bibitem{poten1}{}
N. Barik, and B. K. Dash, Phys. Rev. D {\bf 33}, 1925 (1986).

\bibitem{poten2}{}
T. Fredericot, B. V. Carlson, R. A. R\^{e}go and M. S. Hussein, J. Phys. G: Nucl. Part. Phys. {\bf 15}, 297 (1989).

\bibitem{poten3}{}
E. F. Batista, B. V. Carlson, and T. Frederico, Nucl. Phys. A {\bf 697}, 469 (2002).

\bibitem{panda1}{}
N. Barik, R. N. Mishra, D. K. Mohanty, P. K. Panda, and T. Frederico, Phys. Rev. C {\bf 88}, 015206 (2013).

\bibitem{panda2}{}
R. N. Mishra, H. S. Sahoo, P. K. Panda, N. Barik, and T. Frederico, Phys. Rev. C {\bf 92}, 045203 (2015).

\bibitem{panda3}{}
R. N. Mishra, H. S. Sahoo, P. K. Panda, N. Barik, and T. Frederico, Phys. Rev. C {\bf 94}, 035805 (2016).

\bibitem{humq1}{}
X. Y. Xing, J. N. Hu, and H. Shen, Phys. Rev. C {\bf 94}, 044308 (2016).

\bibitem{humq2}{}
X. Y. Xing, J. N. Hu, and H. Shen, Phys. Rev. C {\bf 95}, 054310 (2017).

\bibitem{humq3}{}
J. N. Hu, and H. Shen, Phys. Rev. C {\bf 96}, 054304 (2017).

\bibitem{zhu18}{}
Z.-Y. Zhu, E.-P. Zhou, and A. Li, Astrophys. J. {\bf862}, 98 (2018).

\bibitem{0348+0432}{}
J. Antoniadis, P. C. C. Freire, N. Wex, T. M. Tauris, R. S. Lynch, M. H. van Kerkwijk, M. Kramer, C. Bassa, V. S. Dhillon, T. Driebe, J. W. T. Hessels, V. M. Kaspi, V. I. Kondratiev, N. Langer, T. R. Marsh, M. A. McLaughlin, T. T. Pennucci, S. M. Ransom, I. H. Stairs, J. van Leeuwen, J. P. W. Verbiest, and D. G. Whelan, Science {\bf 340}, 1233232 (2013).

\bibitem{1614-2230-1}{}
P. B. Demorest, T. Pennucci, S. M. Ransom, M. S. E. Roberts, and J. W. T. Hessels, Nature {\bf 467}, 1081 (2010).

\bibitem{1614-2230-2}{}
E. Fonseca, T. T. Pennucci, J. A. Ellis, I. H. Stairs, D. J. Nice, S. M. Ransom, P. B. Demorest, Z. Arzoumanian, K. Crowter, T. Dolch, R. D. Ferdman, M. E. Gonzalez, G. Jones, M. L. Jones, M. T. Lam, L. Levin, M. A. McLaughlin, Kevin Stovall, J. K. Swiggum, and W. Zhu, Astrophys. J. {\bf 832}, 167 (2016).

\bibitem{baoan}{}
B. A. Li, and X. Han, Phys. Lett. B {\bf 727}, 276 (2013).

\bibitem{steiner}{}
A. W. Steiner, M. Hempel, and T. Fischer, Astrophys. J. {\bf 774}, 17 (2013).

\bibitem{HICsci}{}
P. Danielewicz, R. Lacey, and W. G. Lynch, Science {\bf 298}, 1592 (2002).

\bibitem{IAS}{}
P. Danielewicza, and J. Lee, Nucl. Phys. A {\bf 922}, 1 (2014).

\bibitem{alphaD}{}
Z. Zhang, and L. W Chen, Phys. Rev. C {\bf 92}, 031301 (2015).

\bibitem{HICsym}{}
M. B. Tsang, Y. Zhang, P. Danielewicz, M. Famiano, Z. Li, W. G. Lynch, and A. W. Steiner, Phys. Rev. Lett. {\bf 102}, 122701 (2009).

\bibitem{Elias}{}
E. R. Most, L. R. Weih, L. Rezzolla, and J. Schaffner-Bielich, Phys. Rev. Lett. {\bf 120}, 261103 (2018).

\bibitem{Fattoyev}{} 
F.~J. Fattoyev, J. Piekarewicz, and C.~J. Horowitz, Phys. Rev. Lett. {\bf 120}, 172702 (2018).

\bibitem{Annala}{} 
E. Annala, T. Gorda, A. Kurkela, and A. Vuorinen, Phys. Rev. Lett. {\bf 120}, 172703 (2018).

\bibitem{Bauswein}{}
A. Bauswein, O. Just, H. Janka, and N. Stergioulas, Astrophys. J. Lett. {\bf 850}, L34 (2017).

\bibitem{lvsr1}{}
J. M. Lattimer, and M. Prakash, Astrophys. J. {\bf 550}, 426 (2001).

\bibitem{lvsr2}{}
J. M. Lattimer, and M. Prakash, Science {\bf 304}, 536 (2004).

\bibitem{lvsr3}{}
B.-A. Li, and A. W. Steiner, Phys. Lett. B {\bf 642}, 436 (2006).


\end{thebibliography}
\end{document}